\documentstyle[prb,aps,multicol,epsf]{revtex}
\begin{document}
\draft
\title{Critical behavior of the 3d random field Ising model:\\ 
Two-exponent scaling or first order phase transition?}
\author{Heiko Rieger}
\address{
Institut f\"ur Theoretische Physik, Universit\"at zu K\"oln, 50926 K\"oln,Germany\\
and\\
HLRZ c/o Forschungszentrum J\"ulich, Postfach 1913, 
52425 J\"ulich, Germany}

\date{March 7, 1995}
\maketitle
\begin{abstract}
In extensive Monte Carlo simulations the phase transition of the
random field Ising model in three dimensions is investigated. The
values of the critical exponents are determined via finite size
scaling.  For a Gaussian distribution of the random fields it is found
that the correlation length $\xi$ diverges with an exponent
$\nu=1.1\pm0.2$ at the critical temperature and that
$\chi\sim\xi^{2-\eta}$ with $\eta=0.50\pm0.05$ for the connected
susceptibility and $\chi_{\rm dis}\sim\xi^{4-\overline{\eta}}$ with
$\overline{\eta}=1.03\pm0.05$ for the disconnected susceptibility.
Together with the amplitude ratio $A=\lim_{T\to T_c}\chi_{\rm
dis}/\chi^2(h_r/T)^2$ being close to one this gives further support
for a two exponent scaling scenario implying $\overline{\eta}=2\eta$.
The magnetization behaves discontinuously at the transition, i.e.\
$\beta=0$, indicating a first order transition. However, no divergence
for the specific heat and in particular no latent heat is found. Also
the probability distribution of the magnetization does not show a
multi-peak structure that is characteristic for the phase-coexistence at
first order phase transition points.

\end{abstract}
\pacs{75.10.H, 05.50, 64.60.C}

\begin{multicols}{2}

\section{Introduction}

The critical behavior of the random field Ising model (RFIM) has been
investigated intensively over two decades now (see \cite{Review} for a
review) but a full agreement over the most fundamental issues still
missing. Although it has been shown rigorously \cite{Bricmont} that at
low temperatures and sufficiently small field strength there is indeed
long range ferromagnetic order in three dimensions, the
characterization of the phase transition is still unclear. The idea of
dimensional reduction \cite{Dim_red} failed to describe the critical
behavior of the model correctly. The reason for this was essentially
the presence of many local minima in the free energy landscape that do
not allow for a straightforward perturbative treatment. A reminiscence
of dimensional reduction is still present in the scaling theories for
the RFIM \cite{Villain,Fisher,Bray}, where the dimension $d$ has been
replaced by $d-\theta$ in hyper-scaling relations and $\theta$ was
introduced as a third exponent independent of $\nu$ and $\eta$.
However, these (droplet)-theories relied upon the assumption of a
second order phase transition. Part of the results obtained via Monte
Carlo simulations support this scenario \cite{Og,OgHu}, whereas other
simulations find indications for the transition to be of first order
\cite{Young,Rieger}. More surprisingly no latent heat was found at the
transition \cite{Rieger}, contrary to what one might expect at a first
order phase transition \cite{Nauenberg}, where a discontinuity in the
internal energy at the critical point causes the specific heat to
diverge with the volume of the system. Actually no divergence at all
could be detected for the specific heat \cite{Rieger}.

The experimental situation is also far from clear. Experiments on
diluted antiferromagnets in an uniform external field (DAFF), which
have been demonstrated to be in the same universality class as the
RFIM \cite{Fishman,Cardy}, provided evidence for a second order phase
transition with varying estimates for the critical exponents (for a
discussion see \cite{Review}). In \cite{Belanger85} a logarithmic
singularity for the specific heat was found, implying
$\alpha=0$. Quite recently, subsequent to the prediction made in
\cite{Rieger}, experimental evidence of a cusp-like singularity of the
specific heat, i.e.\ $\alpha<0$, was reported \cite{Karszewiski}.
Moreover, other possible scenarios, like a first order transition
\cite{Birgeneau} or a so called ``trompe l'oeil critical behavior''
\cite{Hill} have also been discussed. Experiments addressing the
equilibrium quantities of the RFIM (or the DAFF, respectively) are
extremely difficult to perform, since by approaching the transition
the system falls out of equilibrium very rapidly and freezes into a
metastable domain state (see \cite{Kleemann} for a review). Therefore
these issues have not been settled by now.

The purpose of this paper is to estimate {\it all} critical exponents
with the help of extensive Monte Carlo investigations of the RFIM in
three dimensions. In a recent paper results for the exponents of the
RFIM with a binary distribution were presented \cite{Rieger}. In mean
field theory \cite{Aharony} there is a sharp discrepancy between the
phase diagram for a binary distribution and that of a continuous
distribution with a finite probability for zero random fields: in the
former case the phase diagram has a tritcritical point, where the
phase transition changes from second order for low fields to first
order at higher fields; in the latter case the transition is always of
second order. One might speculate that a similar scenario occurs also
in three dimensions and that the indications found in \cite{Rieger}
for a first order transition are in fact an artifact of the binary
distribution. Therefore it is important to perform a similar analysis
for a continuous distribution, which is presented in this paper.

It is organized as follows: In section 2 we present the model and
review various scaling predictions that are supposed to hold in case
he transition is of second order. Section 3 reports our results of the
finite size scaling analysis and section 4 discusses the possibility
of a first order transition. Section 5 is a summarize and an outlook.

\section{Model and Scaling Theory}

The random field Ising model (RFIM) in three dimensions is defined by
the Hamiltonian
\begin{equation}
H= - J \sum_{\langle i,j\rangle} S_i S_j - \sum_i h_i S_i\;,
\label{model}
\end{equation}
where $S_i=\pm1$ are Ising spins, the first sum is over nearest
neighbor pairs on a simple cubic lattice and the $h_i$ are independent
quenched random variables with mean zero $[h_i]_{\rm av}=0$ and
variance $[h_i^2]_{\rm av}=h_r^2$. It has been shown rigorously
\cite{Bricmont} that this model has a transition to long range
ferromagnetic order along a line $h_r(T)$ in the $(h_r,T)$-diagram. 

If this transition is second order, a scaling theory has been proposed
\cite{Villain,Fisher,Bray} that relies upon the assumption that random
field induced fluctuations dominate over thermal fluctuations at
$T_c$. This implies for the singular part of the free energy
\begin{equation}
F_{\rm sing}\sim\xi^{\theta}\;,         \label{fsing}
\end{equation}
where $\xi$ is the correlation length diverging at the critical temperature
like 
\begin{equation}
\xi\sim t^{-\nu}\quad{\rm with}\quad t=T-T_c
\end{equation}
and $\theta$ is a new exponent.
Random field fluctuations alone produce typically an excess field of the order
of $\xi^{d/2}$ within a correlation volume, so a naive guess would be
$\theta$ roughly 1.5 in three dimensions. From (\ref{fsing}) the
modified hyper-scaling relation follows
\begin{equation}
2-\alpha=\nu(d-\theta)\;,               \label{hyper}
\end{equation}
$\alpha$ being the specific heat exponent. It also implies an
exponential divergence of the relaxation time $\tau$ at $T_c$:
$\tau\sim\exp(A\xi^\theta)$. The decay of the connected and
disconnected correlation functions at $T_c$
\begin{eqnarray}
C({\bf r}) = & [\langle S_0 S_r\rangle
- \langle S_0\rangle\langle S_r\rangle]_{\rm av}
& \sim r^{-(d-2+\eta)}                \label{corr}\\
C_{\rm dis} ({\bf r}) = &
\lbrack\langle S_0 \rangle\langle S_r \rangle\rbrack_{\rm av}
& \sim r^{-(d-4+\overline{\eta})}     \label{corrdis}
\end{eqnarray}
defines two exponents $\eta$ and $\overline{\eta}$. These are expected
\cite{Villain,Fisher} to be related to the new exponent $\theta$ via
\begin{equation}
\theta=2-\overline{\eta}+\eta\;.        \label{theta}
\end{equation}
By approaching the critical temperature from above the corresponding
connected and disconnected susceptibilities diverge as
\begin{eqnarray}
\chi = & \int d{\bf r}\; C({\bf r})
& \sim (T-T_c)^{-\gamma}\;,            \label{sus}\\
\chi_{\rm dis} = & \int d{\bf r}\; C_{\rm dis}({\bf r})
& \sim (T-T_c)^{-\overline{\gamma}}\;. \label{susdis}
\end{eqnarray}
These exponents are related by
\begin{eqnarray}
\gamma & = & \nu(2-\eta)\;,             \label{gamma}\\
\overline{\gamma} & = & \nu(4-\overline{\eta})\;.  
                                        \label{gammad}
\end{eqnarray}
In addition one has the usual scaling relations
for the magnetization exponent
\begin{equation}
\beta =  (d-4+\overline{\eta})\;,    \label{beta}
\end{equation}
and the Rusbrooke equality
\begin{equation}
\alpha+2\beta+\gamma = 2            \label{Rushbrooke}\;.
\end{equation}
Obviously there seem to be three
independent critical exponents and a central issue of the activities
on the critical properties of the RFIM is the quest for an additional
scaling relation. Already Imry and Ma \cite{Imry} conjectured that
$F_{\rm sing}\sim\chi$, where $\chi\sim (T-T_c)^{-\gamma}$ is the 
susceptibility. This would imply $\theta=2-\eta$ via (\ref{fsing}) and
therefore with (\ref{theta})
\begin{equation}
\overline{\eta}=2\eta\;.                \label{twoexp}
\end{equation}
A set of analytical arguments by Schwartz et al.\
\cite{Schwartz85,Schwartz} supports this two-exponent scaling scenario
indicated by (\ref{twoexp}). Hence the exact Schwartz-Soffer
inequality \cite{Schwartz85} $\overline{\eta}\le2\eta$ might be
fulfilled as an equality, as in (\ref{twoexp}). A recent high
temperature series analysis provided \cite{Gofman} strong evidence of
this equality to hold by estimating $\gamma$ and $\overline{\gamma}$
via Pad\'e-approximation.

In this paper we intend to determine {\it all} critical exponents with Monte
Carlo simulations. The model is defined by the Hamiltonian (\ref{model})
on a finite lattice of linear size $L$ with periodic boundary conditions.
Instead of a binary distribution, as used in \cite{Rieger}, we consider
here a Gaussian distribution 
\begin{equation}
P(h_i) = \frac{1}{\sqrt{2\pi h_r^2}}\,\exp(-h_i^2/2h_r^2)
\label{gauss}
\end{equation}
for the random fields. Unfortunately this makes the use of multispin
coding impossible \cite{Multi}, which slows down the spin-update algorithm
(Metropolis) by approximately a factor 10. The simulations were done
at a constant ratio $h_r/T = 0.35$.

As was pointed out in \cite{Rieger} it is of paramount importance to
perform an extensive disorder average, since quantities like the
susceptibility $\chi$ are non-self-averaging in this model. Therefore
at least 1000 samples for each temperature and system size were
used. Equilibration was checked by starting each sample with two
different initial configurations, one with all spins down and the other
with all spin up, and running as long as both ``replica'' yield the
same values for all thermodynamic quantities of interest. For $L=16$
up to $2\cdot10^6$ Monte Carlo sweeps were necessary to fulfill this
criterion.

The following quantities were calculated for each disorder
realization: The average magnetization per spin $\langle
M\rangle$, its square $\langle M^2\rangle$, the average
energy per spin $\langle E\rangle$ and its square $\langle
E^2\rangle$. The angular brackets, $\langle \ldots \rangle $,
denote a thermal average for a single random field configuration
and $M=1/N\sum_{i=1}^N S_i$, $E=H(\underline{S})/N$ and
$N=L^d$. From these data we get the specific heat per spin, $C$, 
the susceptibility $\chi$, the disconnected susceptibility, 
$\chi_{{\rm dis}}$, and the order parameter, $m$, as follows
\begin{eqnarray}
[C]_{\rm av} & = & N\Bigl\{[\langle E^2\rangle]_{\rm av}
-[\langle E\rangle^2]_{{\rm av}}\Bigr\}/T^2\;,\label{defc}\\
\lbrack m\rbrack_{\rm av}& = & [\vert\langle M\rangle\vert]_{\rm
  av}\;,\label{defm}\\
\,[\chi]_{\rm av} & = & N\Bigl\{[\langle M^2\rangle]_{\rm av}
-[\langle M\rangle^2]_{{\rm av}}\Bigr\}/T\;,\label{defchi}\\
\lbrack\chi_{\rm dis}\rbrack_{\rm av} & = & N[\langle
M\rangle^2]_{{\rm av}}\;,\label{defchid}
\end{eqnarray}
where $[\ldots ]_{{\rm av}}$ denotes the average over different random field
configurations. The finite size scaling function of any quantity $A(T,L)$,
which behaves in the infinite system like $A_\infty\sim t^{-a}$ by
approaching the critical temperature, reads
\begin{equation}
A(T,L) = L^{a/\nu}\tilde{A}(t L^{1/\nu})\;.
\end{equation}
In case of the specific heat the scaling function $\tilde{C}(x)$ has a
maximum at some value $x=x^*$.  For each lattice--size we estimate the
temperature $T^*(L)$, where $T^2[C]_{\rm av}$ is maximal. Since
$t^*(L)\,L^{1/\nu}=x^*$ we obtain in this way the critical temperature
$T_c$ and the correlation length exponent $\nu$ from
\begin{equation}
t^*(L) \equiv T^*(L) - T_c = x^*L^{-1/\nu}\;.     \label{tc}
\end{equation}
We denote the value of $T^*(L)^2 [C]_{\rm av}$
at this temperature $T^*(L)$ by $[C^*]_{\rm av}$ and similarly for the
other quantities in Eq. (\ref{defc}-\ref{defchid}):
$[\chi^*]_{\rm av} \equiv T^*(L)\,[\chi(L,T^*(L))]_{\rm av}$,
$[\chi^*_{\rm dis}]_{\rm av} \equiv [\chi_{\rm dis}(L,T^*(L))]_{\rm av}$
and
$[m^*]_{\rm av} \equiv [m(L,T^*(L))]_{\rm av}$. Then one expects the
following behavior:
\begin{eqnarray}
\,[C^*]_{\rm av} 
& \sim & L^{\alpha/\nu}\,\tilde{C}(x^*)\;,         \label{cmax}\\
\,[\chi^*]_{\rm av} 
& \sim & L^{2-\eta}\,\tilde{\chi}(x^*)\;,            \label{chi}\\
\,[\chi^*_{\rm dis}]_{\rm av}\; 
& \sim & L^{4-\overline{\eta}}\,\tilde{\chi}_{\rm dis}(x^*)\;,
                                                  \label{chid}\\
\,[m^*]_{\rm av}\; 
& \sim & L^{-\beta/\nu}\,\tilde{m}(x^*)\;.           \label{mag}
\end{eqnarray}
Through the size dependence of $[C^*]_{\rm av}$, $[\chi^*]_{\rm av}$,
$[\chi_{\rm dis}^*]_{\rm av}$ and $[m^*]_{\rm av}$ we obtain the
exponents $\alpha$, $\eta$, $\overline{\eta}$ and $\beta$.
In the vicinity of $x^*$ the scaling function $\tilde{C}(x)$ can be
approximated by a parabola. Therefore three temperatures near the
maximum of the specific heat are enough to determine the values of
$T^*(L)$ as well as $[C^*]_{\rm av}$ etc.  Let us assume that we
measured $C_1\pm\Delta_1$, $C_2\pm\Delta_2$ and $C_3\pm\Delta_3$ for
temperatures $T_1<T_2<T_3$. In order to get a reliable estimate for
$\overline{C}(L)$ as well as its errorbar we have to have
$C_1+\Delta_1<C_2-\Delta_2$ and $C_3+\Delta_3<C_2-\Delta_2$, hence it
is important to have very small statistical errors $\Delta_1$,
$\Delta_2$ and $\Delta_3$. The difference between the temperatures
should be small, otherwise the second order polynomial approximation
fails, but not too small, since then the above requirement for the
difference between $C_1$, $C_2$ and $C_3$ cannot be fulfilled. We used
$T_2-T_1=T_3-T_2=0.1$ for the small lattice sizes ($L<10$) and $0.05$
for the larger sizes.  Once $T^*(L)$ is determined, the value of the
quantities in (\ref{cmax}-\ref{mag}) are again determined by fitting a
second order polynomial. Finally, the exponents are determined via
least square fits of the data obtained to the functional forms given
in (\ref{cmax}-\ref{mag}).

\section{The critical exponents}

\begin{figure}
\epsfxsize=10cm
\epsffile{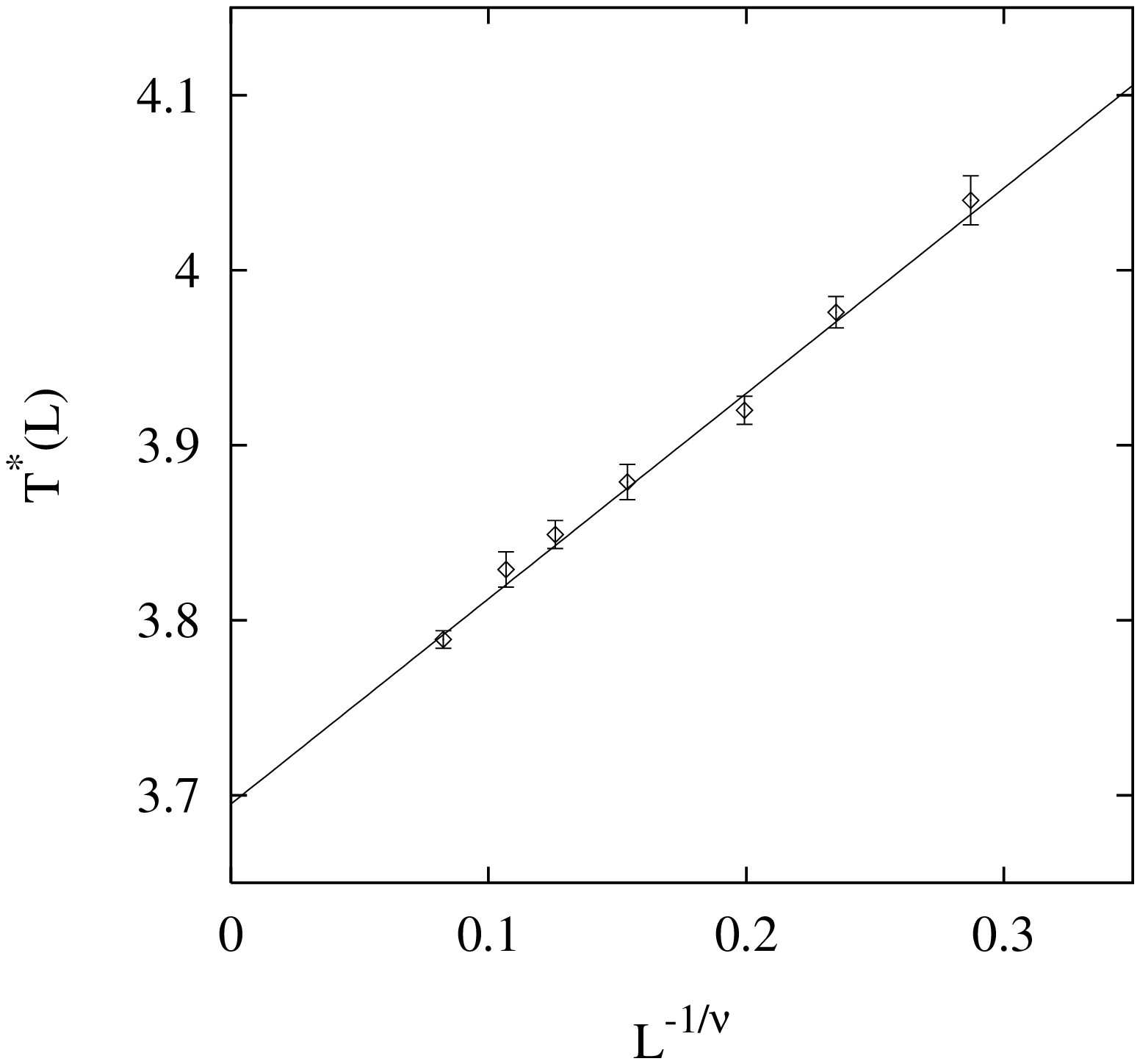}
\vskip0.9cm
\caption{ \label{fig1}
The temperature $T^*(L)$, where the specific heat attains its 
maximum, versus $L^{1/nu}$, with $\nu$ as in (\protect{\ref{nufit}}).
From right to left one has $L=4$, $5$, $6$, $8$, $10$, $12$ and $16$.
The full line is a least square straight line fit.}
\end{figure}

First we determine the temperature $T^*(L)$, where the maximum of the
specific heat occurs and fit the correlation length exponent
$\nu$ in such a way that the data for $T^*(L)$ lie on a straight line
if plotted against $L^{-1/\nu}$. Then relation (\ref{tc}) is 
fulfilled and $T_c$ can be read of from the intersection of this line with
the $y$-axis. The result is depicted in figure (\ref{fig1}), one obtains
\begin{equation}
\nu=1.1\pm0.2\quad{\rm and}\quad T_c = 3.695\pm0.02\;.
\label{nufit}
\end{equation}

\begin{figure}
\epsfxsize=10cm
\epsffile{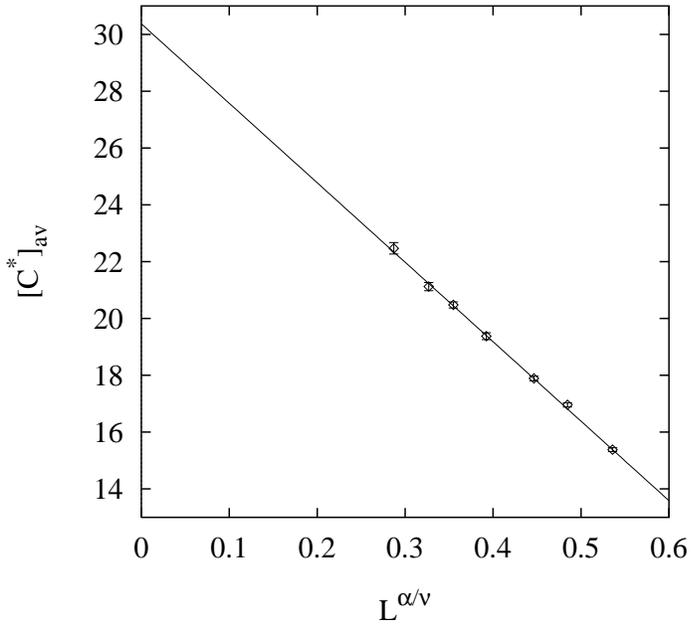}
\vskip0.9cm
\caption{ \label{fig2}
The maximum of the specific heat $[C^*]_{\rm av}$ versus 
$L^{\alpha/\nu}$ with $\alpha/\nu$ as in (\protect{\ref{nufit}}).
From right to left one has $L=4$, $5$, $6$, $8$, $10$, $12$ and $16$.
The full line is a least square straight line fit.}
\end{figure}

Next we analyse the size-dependence of maximum of the specific heat.
According to (\ref{cmax}) one should be able to read off the
exponent-ratio $\alpha/\nu$ from the slope of a straight line fit in a
log-log plot of $[C^*]_{\rm av}$ versus $L$. The first observation is
that this procedure does not work since the data points show a
significant negative curvature in a log-log plot, indicating a much
weaker dependency then algebraic. In reminiscence of the experimental
finding of a {\it logarithmic} divergence of the specific heat
(i.e.\ $\alpha=0$) we plotted $[C^*]_{\rm av}$ versus $\log(L)$,
and still found a (now less pronounced) negative curvature in
the data points, indicating a still weaker or even no divergence. 
Finally we hypothesized a cusp-like singularity, i.e.\ 
\begin{equation}
c_\infty - [C^*]_{\rm av} \sim L^{\alpha/\nu}
\end{equation}
with $\alpha<0$. The result of the corresponding fitting procedure 
is shown in figure \ref{fig2}. As one can see the data are fully
consistent with this type of dependency and we get
\begin{equation}
\alpha/\nu=-0.45\pm0.05\quad{\rm and}\quad c_\infty=30.4\pm0.1.
\label{alphafit}
\end{equation}

With the estimate for $\nu$ from (\ref{nufit}) it is $\alpha=-0.5\pm0.2$.
Although the ratio $\alpha/\nu$ can be determined quite
accurately, the value for $\alpha$ alone depends on $\nu$ which
has a much larger large error bar.

\begin{figure}
\epsfxsize=10cm
\epsffile{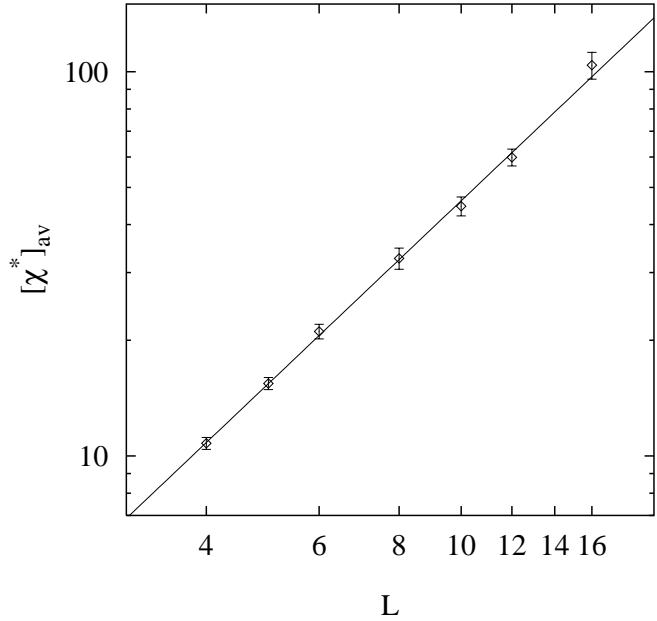}
\vskip0.9cm
\caption{ \label{fig3}
The value of the susceptibility $[\chi]_{\rm av}$ at the temperature
$T^*(L)$ versus system size $L$ in a log-log plot.
The full line is a least square straight line fit, which
gives a slope of $1.55\pm0.05$  that is an estimate for $2-\eta$
according to (\protect{\ref{chi}}).}
\end{figure}

According to (\ref{chi}) the susceptibility at the temperature
$T^*(L)$ should diverge algebraically, and the exponent $\eta$
is obtained by a least square straight line fit of
$[\chi^*]_{\rm av}$ versus $L$ in a log-log plot, which is depicted
in figure \ref{fig3}. The result is
\begin{equation}
\eta=0.50\pm0.05\;,\label{etafit}
\end{equation}
from which one obtains via (\ref{gamma}) $\gamma=1.7\pm0.2$.

\begin{figure}
\epsfxsize=10cm
\epsffile{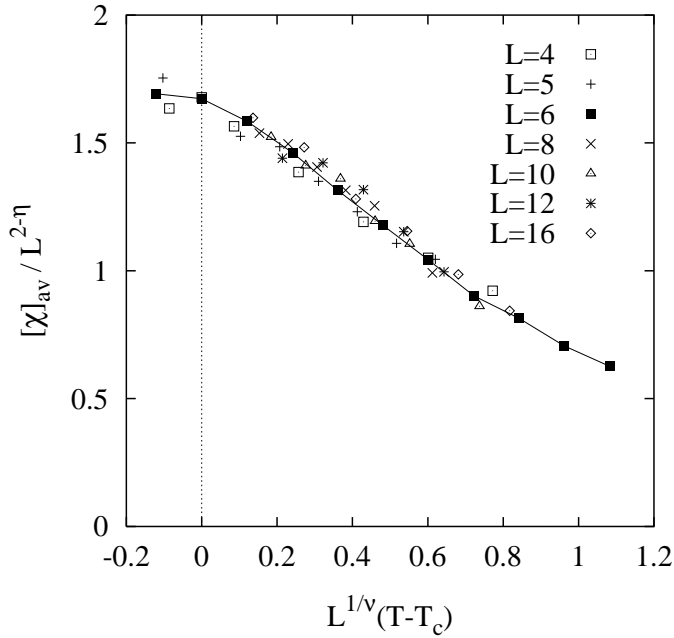}
\vskip0.9cm
\caption{ \label{fig4}
A scaling plot of the susceptibility $[\chi]_{\rm av}$ with the parameters
for $T_c$, $\nu$ and $\eta$ given in (\protect{\ref{tc}})
and (\protect{\ref{chi}}).The full line connects only the points for $L=6$
as a guide for the eyes.}
\end{figure}

The finite size scaling behavior of $\chi$ near the critical
temperature is expected to be
\begin{equation}
[\chi]_{\rm av}(T,L)
\approx L^{2-\eta}\tilde{\chi}\bigl(L^{1/\nu}(T-T_c)\bigr)\;.
\end{equation}
In figure \ref{fig4} the corresponding finite size scaling plot 
using the values for $T_c$, $\nu$ and $\eta$ from (\ref{nufit})
and (\ref{etafit}) is shown.

\begin{figure}
\epsfxsize=10cm
\epsffile{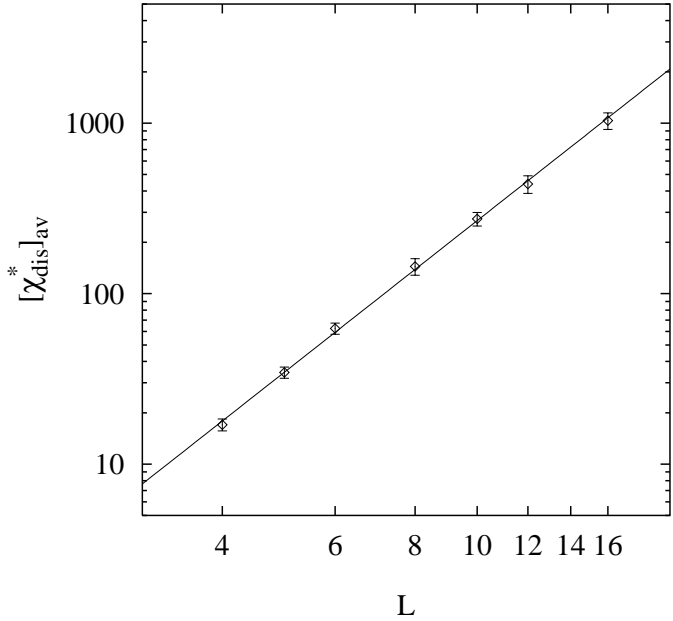}
\vskip0.9cm
\caption{ \label{fig5}
The disconnected susceptibility $[\chi_{\rm dis}]_{\rm av}$ 
at the temperature $T^*(L)$ versus $L$ in a log-log plot. The
full line is a least square straight line fit, which gives a slope of
$2.97\pm0.08$ that is an estimate for $4-\overline{\eta}$ according to
(\protect{\ref{chid}}).}
\end{figure}

The disconnected susceptibility at $T^*(L)$ is depicted in figure
\ref{fig5}, which should diverge according to (\ref{chid}) with
system size. Via a least square fit of the data to a straight line
in the log-log plot one obtains the estimate
\begin{equation}
\overline{\eta}=1.03\pm0.05\;.\label{etabfit}
\end{equation}

With the scaling relation (\ref{theta}) one can
estimate $\theta$ to be 
\begin{equation}
\theta=1.5\pm0.1\;,\label{thetafit}
\end{equation}
which is in good agreement with the value $\theta=d/2$ mentioned 
in the last section.

Comparing our values for $\eta$ (\ref{etafit}) and $\overline{\eta}$
(\ref{etabfit}) one notes that the relation (\ref{twoexp}) might be
fulfilled, which would support the two-exponent scaling scenario
mentioned in the last section. In order to check this we calculated
the amplitude ratio
\begin{equation}
{\bf A}=
\lim_{T\to T_c}\frac{[\chi_{\rm dis}]_{\rm av}}
{[\chi]_{\rm av}^2(h_r/T)^2}\;.
\label{amp}
\end{equation}
as suggested in \cite{Gofman}, where in an extensive high-temperature
analysis it was found that $A=1$, implying strong evidence for
two-exponent scaling.
The finite size scaling form of (\ref{amp}) is
${\bf A}(L,T)=\tilde{\bf A}(L^{1/\nu}(T-T_c))$ and therefore we expect
\begin{equation}
{\bf A}^*(L)\equiv{\bf A}\bigl(L,T^*(L)\bigr)\,=\,1
\end{equation}
independent of system size. The result that we obtain from our
MC-data is table 1:
\begin{center}
\begin{tabular}{|crc|clc|clc|}
\hline 
$\quad$ &L& $\quad$ & $\quad$ & $A^*$ & $\quad$ & $\quad$ & $\Delta A^*$ & $\quad$\\
\hline\hline
 & 4 &  & & 1.19 & &  & 0.22 & \\
 & 6 &  & & 1.14 & &  & 0.23 & \\
 & 8 &  & & 1.10 & &  & 0.33 & \\
 & 10 & & & 1.12 & &  & 0.29 & \\
 & 12 & & & 1.00 & &  & 0.26 & \\
 & 16 & & & 0.80 & &  & 0.27 & \\
\hline
\end{tabular}
\end{center}
\vskip0.2cm
{\small TABLE I: The amplitude ratio $A^*(L)=A\bigl(L,T^*(L)\bigr)$
and its estimated error for different system sizes.}
\vskip0.3cm

As one can see $A^*$ is close to one, which supports the
equality (\ref{twoexp}). Because of the large errorbars
the slight decrease in the estimate of $A^*$ with increasing
system size seems not to be significant.

\begin{figure}
\epsfxsize=10cm
\epsffile{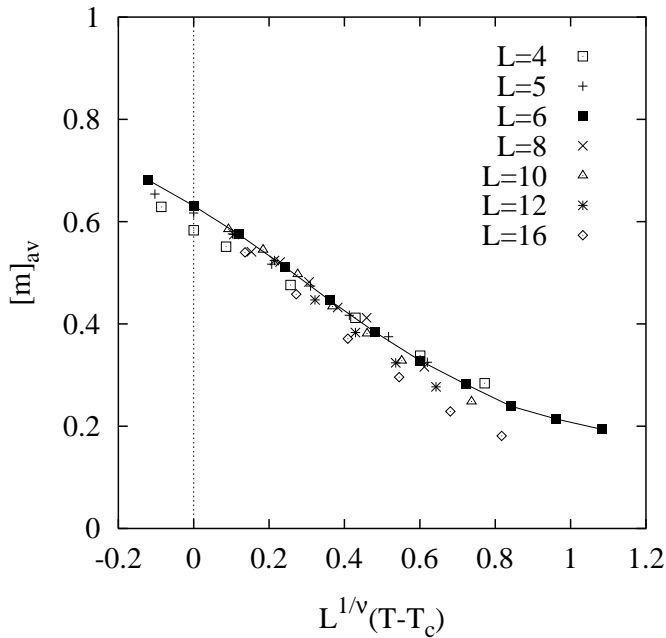}
\vskip0.9cm
\caption{ \label{fig6}
A scaling plot of the magnetization $[M]_{\rm av}$ with the parameters
for $T_c$ and $\nu$ given in (\protect{\ref{tc}}).
The full line connects only the points for $L=6$
as a guide for the eyes. Note that the data for $[M]_{\rm av}$
are {\it not} rescaled by a factor $L^{-\beta/\nu}$, implying
that indeed $\beta=0$.}
\end{figure}

From equation (\ref{etabfit}) one concludes that the disconnected
susceptibility diverges roughly with the volume of the system, since
from (\ref{gammad}) one obtains
$\overline{\gamma}/\nu=4-\overline{\eta}\approx3=d$. If one looks at
equation (\ref{defchid}), where $\chi_{\rm dis}$ is defined, this
implies that $[\langle M\rangle^2]_{\rm av}=\chi_{\rm dis}/L^d$ is
then independent of system size at the critical temperature. Since the
magnetization is defined in a similar manner (\ref{defm}), namely the
square replaced by the modulus, it does not come as a surprise that
one observes that also the magnetization is also independent of system
size at the critical point. This connection manifests itself in the
relation (\ref{beta}), which predicts $\beta\approx0$ if
$\overline{\eta}\approx1$. By looking at $[m^*]_{\rm av}$ as a
function of $L$ we cannot detect any significant variation with system
size. As an even stronger evidence we show in figure
\ref{fig6} a finite size scaling plot of the magnetization according
to
\begin{equation}
[m]_{\rm av}(T,L)\approx
\tilde{m}\bigl(L^{1/\nu}(T-T_c)\bigr)\quad{\rm if}\quad\beta=0\;,
\end{equation}
using the values for $T_c$ and $\nu$ estimated above. Also by studying
the probability distribution of the magnetization in the next section,
we see that that the system in the thermodynamic limit is already
ordered {\it at} $T_c$. Finally, if the transition is first order,
one expects $\eta=1/2$ \cite{Dayan}, which is in agreement with
our estimate (\ref{etafit}).

\begin{figure}
\epsfxsize=10cm
\epsffile{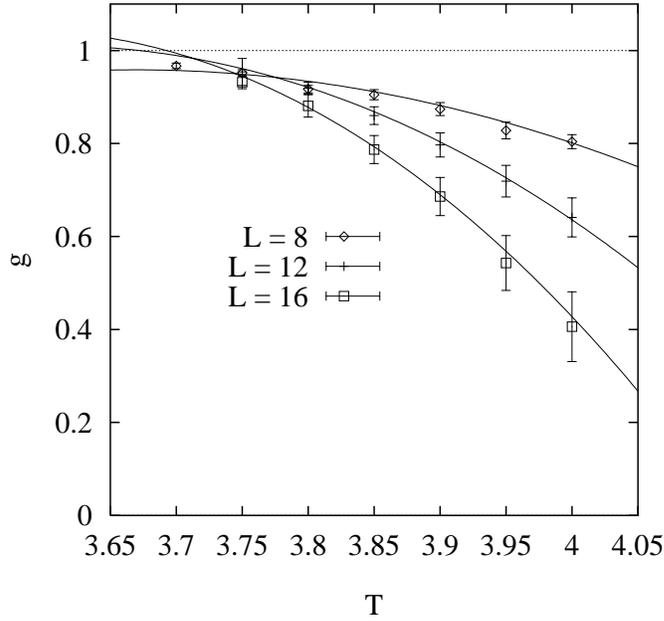}
\vskip0.9cm
\caption{ \label{fig7}
The cumulant $g$ as defined in (\protect{\ref{cum}}) versus temperature
for three different system sizes. The full curves are least square fits
of third order polynomials to the data points. The intersection of the
$L=16$ and $L=12$ curves lies already close to the estimated critical
temperature $T_c=3.695$ (\protect{\ref{tc}}).}
\end{figure}

Finally we study the dimensionless coupling constant or
Binder cumulant
\begin{equation}
g(T,L)=0.5\left\{3-\frac{[\langle M^4\rangle]_{\rm av}}
{[\langle M^2\rangle^2]_{\rm av}}\right\}\;,
\label{cum}
\end{equation}
which is expected to be independent of size at $T_c$. In figure 
\ref{fig7} we show the result for three different system sizes
as a function of temperature. The intersection of the curves 
of the two largest system sizes lie very close to the estimated value
of $T_c=3.695$. 

One observes that close to the critical temperature the value of
the cumulant $g$ lies already very close to $1$. This means that
the probability distribution for the order parameter is
centered around a non-vanishing value for the magnetization,
implying a large degree of magnetic order already {\it at} $T_c$.
This is fully compatible with $\beta=0$ and a discontinuous jump
in the expectation value for the disorder-averaged magnetization
at the critical temperature

\section{Probability distributions}

\begin{figure}
\epsfxsize=10cm
\epsffile{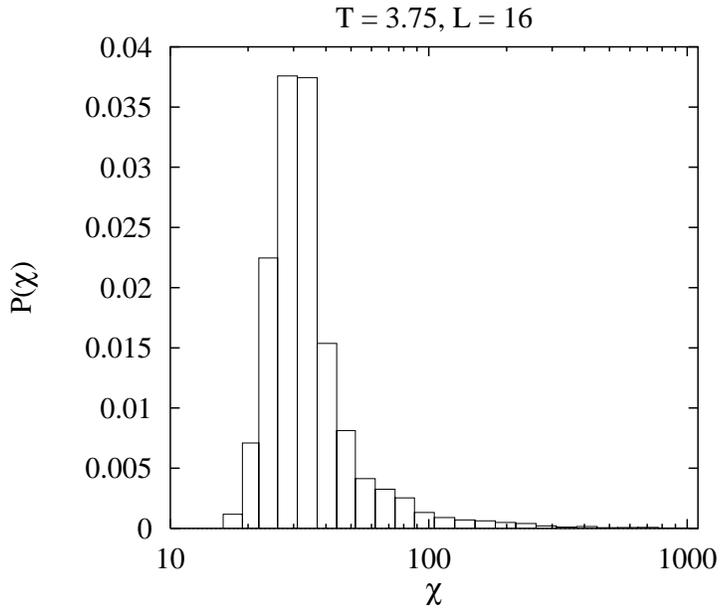}
\vskip0.9cm
\caption{ \label{fig8}
The probability distribution $P(\chi)$ for $L=16$ at the temperature
$T=3.75\approx T^*(L=16)$. Note the logarithmic scale of the abscissa.}
\end{figure}

As mentioned above the susceptibility is a highly non-selfaveraging
quantity, which can be demonstrated by looking at the distribution
$P(\chi)$ for the probability to find a value $\chi$ for the
susceptibility of one sample (i.e.\ realization of the
disorder). $P(\chi)$ is far from being a Gaussian distribution 
and it possesses pronounced long tails as shown in figure \ref{fig8}
(see also \cite{Dayan}).

This is one of the predictions of the droplet theory
\cite{Villain,Fisher}: With a probability proportional to
$L^{-\theta}$ a sample has a vanishing thermodynamic expectation value
for the magnetization leading to a strongly enhanced susceptibility
that is in this case equal to the disconnected susceptibility, as can
be seen from its definition (\ref{defchi},\ref{defchid}). This
argument leads to the equation (\ref{theta}). Moreover it says that
the second moment $[\chi^2]_{\rm av}$ of the distribution $P(\chi)$ is
dominated by these rare events occurring with probability $L^{-\theta}$
and giving a contribution proportional to the square of the
disconnected susceptibility, i.e.\ $(L^{4-\overline{\eta}})^2$. This gives
\begin{equation}
  [\chi^2]_{\rm av}\bigl(L,T^*(L)\bigr)\sim L^\zeta\;,
\end{equation}
with $\zeta=-\theta + 2(4-\overline{\eta})\approx4.5$ via
(\ref{thetafit}) and (\ref{etabfit}).
This gives a much larger contribution than the square of the mean
value $[\chi]_{\rm av}^2\sim L^{2(2-\eta)}\approx L^{3.0}$. By explicitly
calculating $[\chi^2]_{\rm av}(T^*{L},L)$ with our MC-data
we get an estimate of $\zeta=4.0\pm0.2$, which is indeed 
closer to the above droplet prediction of $4.5$.

\begin{figure}
\epsfxsize=10cm
\epsffile{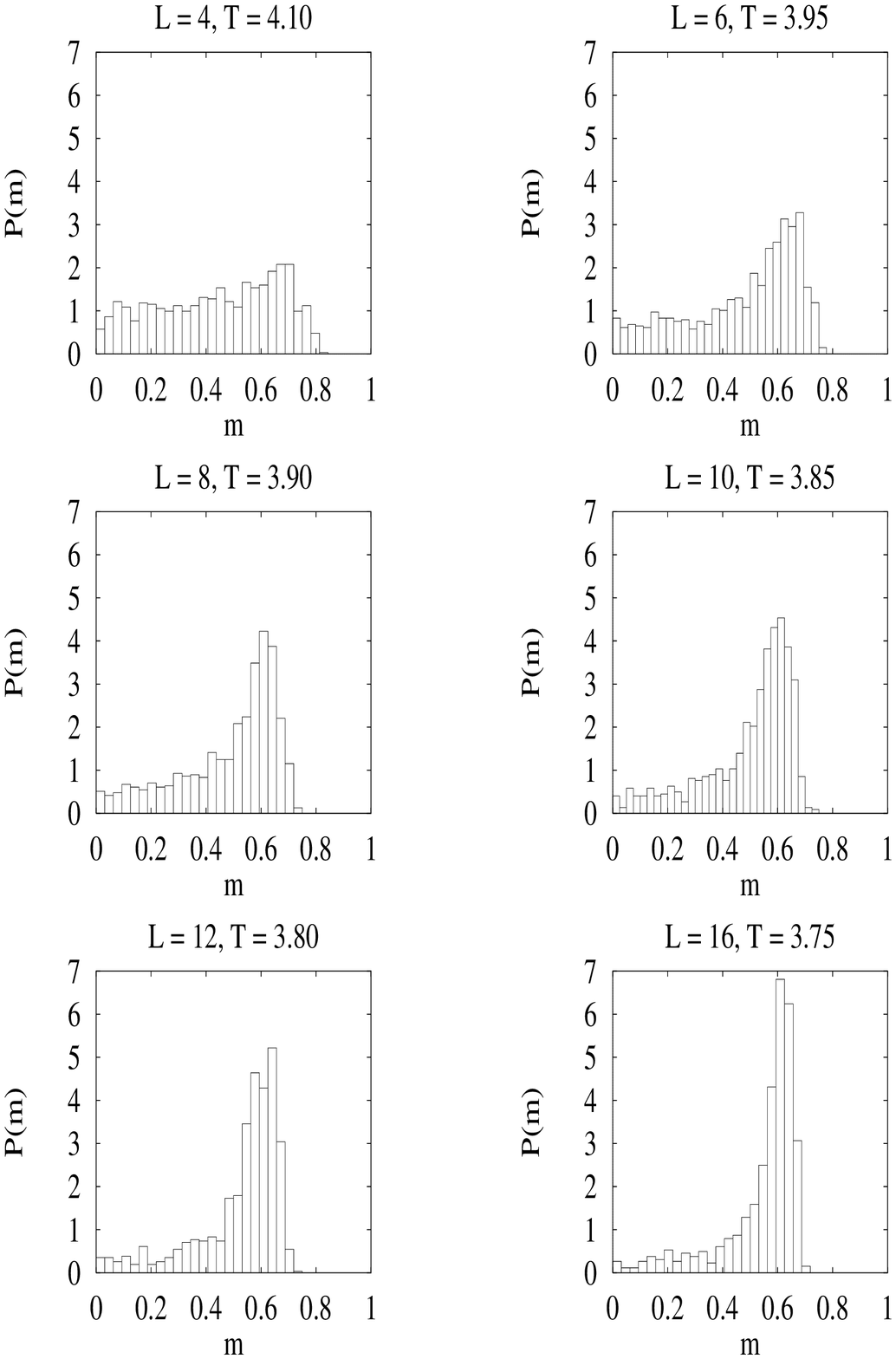}
\vskip0.9cm
\caption{ \label{fig9}
The probability distribution $P_L(m)$ for increasing 
system sizes close to the temperature $T^*(L)$. The tendency
towards a dominant peak at nonvanishing magnetization for
increasing system size is obvious.}
\end{figure}

The scaling hypothesis concerning the probability distribution of the
magnetization can be checked by inspecting the latter directly. In
figure \ref{fig9} we show the probability distribution of the
magnetization for different system sizes at a temperatures close to
$T^*(L)$.  Let us focus at the moment on the behavior of $P_L(m)$ at
zero magnetization. Note that $m$ is defined as the modulus of the
thermodynamic expectation value of the fluctuating quantity $M$ and
can thus take on only positive values.

\begin{figure}
\epsfxsize=10cm
\epsffile{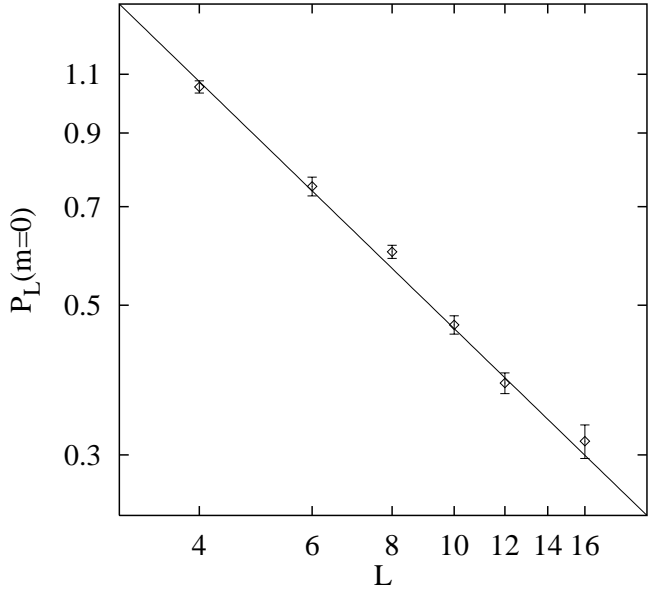}
\vskip0.9cm
\caption{ \label{fig10}
The probability $P_L(m=0,T=T^*(L))$ versus system size $L$
in a log-log plot. The full curve is a least square straight line
fit with slope $\theta=-1.0\pm,0.1$.}
\end{figure}

At a usual second order phase transition
$P_L\bigl(m,T^*(L)\bigr)$ would have a peak at $m=0$ with a width that
is proportional $\chi^*(L)/L^d\sim L^{2-\eta-d}$.  Thus this peak
narrows and gains more weight with increasing system size.  Here it is
quite contrary, the peak develops at a nonzero value $m_\infty$ for
the magnetization.  If it would be a conventional Gaussian
distribution centered around this nonvanishing most probable value for
the magnetization, its value at $m=0$ would decrease exponentially
with system size: $P^*_0(L)\equiv P_L(m=0,T^*(L))\propto
\exp(-m_\infty^2 L^{d-2+\eta})$.  However, according to the above
mentioned scaling theory,
one expects
\begin{equation}
P^*_0(L)\sim L^{-\theta}\;,
\label{p0} 
\end{equation}
which is a much slower decay.
In order to improve the statistics we estimated $P^*_0(L)$ via
$P(m=0)\approx\Delta m^{-1}\,\int_0^{\Delta m} dm\,P(m)$ with $\Delta
m=0.2$ -- $0.3$. In figure \ref{fig10} we depict the result from which
we conclude that (\ref{p0}) obeyed and we estimate $\theta=1.0\pm0.1$,
which is smaller than the value (\ref{thetafit}) obtained by the
relation (\ref{theta}). It is clear that a nonvanishing $\Delta m$
overestimates $P^*_0(L)$ more for larger system sizes, which might 
explain this discrepancy.

Finally we turn our attention to the variation of $P(m)$ with
temperature. According to ref.\ \cite{Binder,Challa} the probability
distribution of the order parameter provides a mean to discriminate
between a first and a second order phase transition. At a usual first
order transition one expects a phase-coexistence at the critical
point, which means that $P(m)$ has two significant peaks, one at zero
(representing the high temperature phase) and one at a positive value
$m_\infty>0$ (representing the low temperature phase). In this case the
effective transition temperature $T_c(L)$ of the finite system can be
identified by the temperature, where both peaks have equal weight
\cite{Challa}.

\begin{figure}
\epsfxsize=10cm
\epsffile{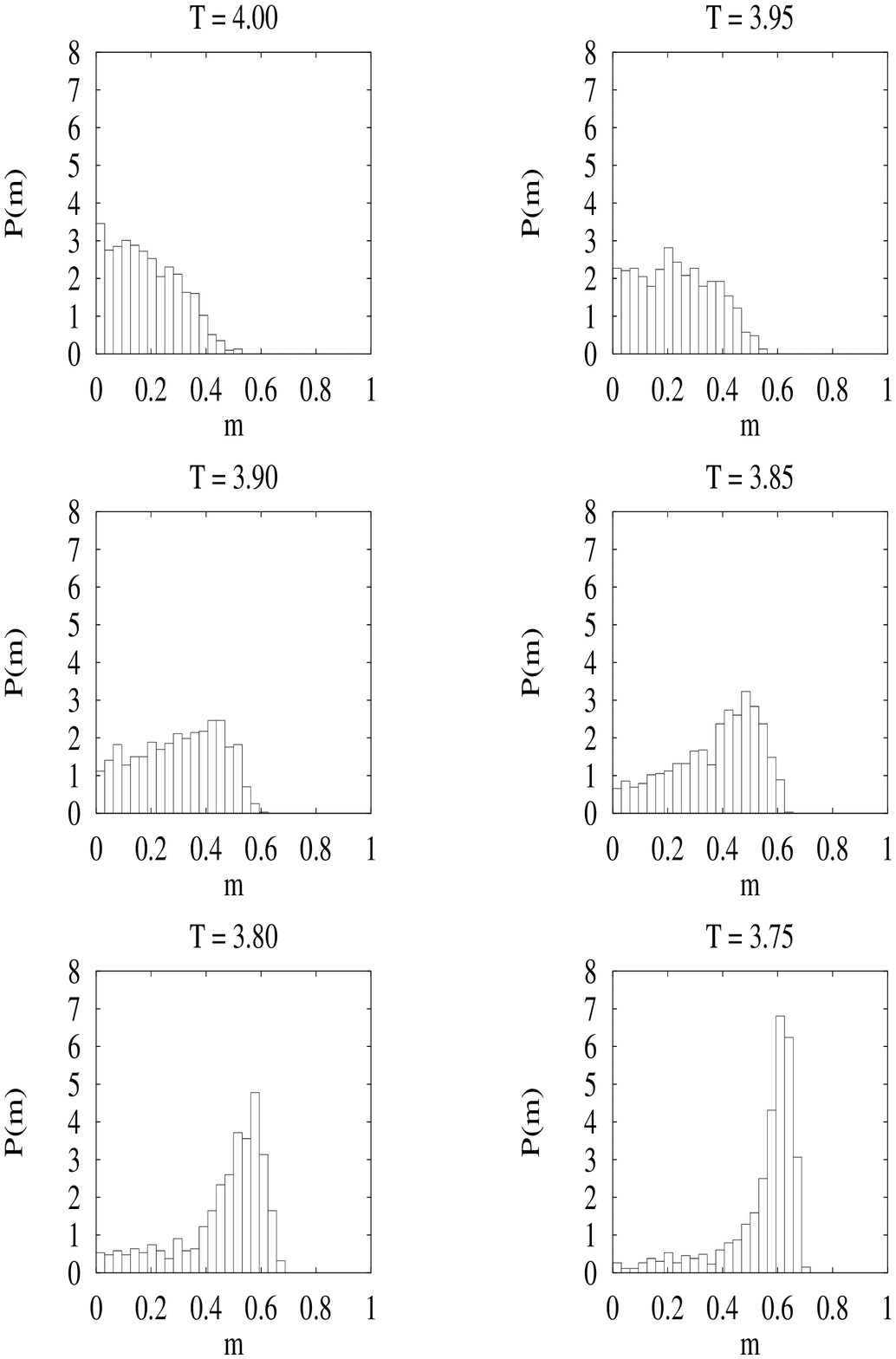}
\vskip0.9cm
\caption{ \label{fig11}
The probability distribution $P(m)$ for $L=16$
at different temperatures. Note that $T=3.80$ is slightly 
above $T^*(L=16)$ and $T=3.75$ is slighly below it, but
still above the estimated value for $T_c=3.695.$}
\end{figure}

If we look at figure \ref{fig11}, where $P_{L=16}(m)$ is shown for
various temperature values, we detect no sign of a characteristic
double-peak structure mentioned above. On the contrary, there is only
a single peak moving continuously to the right with decreasing
temperature. Thus we conclude that phase-coexistence at $T_c$ can be
excluded in the RFIM, although the magnetization jumps discontinuously
there. Hence there is a discontinuity in one of the derivatives of the
free energy (namely with respect to a homogeneous external magnetic field),
which means that the transition is first order in a strict sense.
Nevertheless it is a quite unusual first order phase transition: neither a
latent heat nor a phase-coexistence at $T_c$ could be detected in
our simulations. 

We would like to remark that our findings for the probability
distribution $P_L(m,T)$ are fully compatible with the scenario, which
is suggested by the droplet theory \cite{Villain,Fisher} --- in
particular the systematic decrease of the probability for samples with
zero magnetization (\ref{p0}). Once we accept that $P_L(m=0)$ shrinks
to zero at $T_c$ in the limit $L\to\infty$ the weight of the
distribution has to cumulated at a nonzero most probable value for
$m$. As we have discussed this implies a discontinuous jump of the
average magnetization at $T_c$. Thus one might speculate that our
finding $\beta=0$ is an intrinsic feature of the droplet theory,
although such a possibility has not been discussed yet to our
knowledge.

Another paradigm of usual first order phase transitions is
that the correlation length stays finite at $T_c$, although it might
become very large (for a counter-example see \cite{Anderson}). In this
case the scaling analysis performed in the last section is
unjustified, since it relies on a second order phase transition
scenario with a diverging correlation length.  However, since neither
the specific heat nor the order parameter probability distribution
behaves as usual at a first order transition there seems also to be no
reason us for to expect a finite correlation length at $T_c$.
Finally let us mention that we also looked at the probability
distribution of the internal energy. Here no remarkable behavior
could be observed.

\section{Discussion}

To summarize we list the values for the critical exponents that we
have calculated in table II together with those for the binary
distribution obtained in \cite{Rieger}.
Let us compare these numbers with other known values from the
literature: $\eta=0.25\pm0.03$, $\overline{\eta}\sim0.8$,
$\gamma=1.7\pm0.2$ and indications of a first order transition from
MC-simulations of the RFIM \cite{Young}; $\eta=0.5\pm0.1$ and
$\overline{\eta}=1.0\pm0.3$ from MC-simulations of the DAFF
\cite{OgHu}; $\overline{\eta}=1.1\pm0.1$ and $\nu=1.0\pm0.1$ from a
ground-state investigation via combinatorial optimization \cite{Og};
$\gamma=1.9-2.2$ from real space renormalization group calculations
\cite{Dayan}; $\gamma=2.1\pm0.2$, $\overline{\gamma}=4.2\pm0.4$ from
high temperature series expansion \cite{Gofman}; $\nu=1.25\pm0.11$,
$\overline{\eta}=0.89\pm0.10$ and $\eta=0.4\pm0.8$ from a weighted
mean field theory \cite{Lancaster}. Our estimates are well
compatible with all of these results.

\begin{center}
\begin{tabular}{|clc||clc|l||clc|l|}
\hline 
$\quad$ & & $\quad$ & $\quad$ & Gaussian & $\quad$ & & $\quad$ & Binary & $\quad$ & \\
\hline\hline
&$\eta$            &&& 0.50 && $\pm$0.05 && 0.56 && $\pm$0.03\\
&$\overline{\eta}$ &&& 1.03 && $\pm$0.05 && 1.00 && $\pm$0.06\\
&$\theta$          &&& 1.53 && $\pm$0.1  && 1.56 && $\pm$0.1\\
&$\nu$             &&& 1.1  && $\pm$0.2  && 1.6  && $\pm$0.3\\
&$\gamma$          &&& 1.7  && $\pm$0.2  && 2.3  && $\pm$0.3\\
&$\overline{\gamma}$ &&& 3.3 && $\pm$0.6 && 4.8  && $\pm$0.9\\
&$\beta$           &&& 0.00 && $\pm$0.05 && 0.00 && $\pm$0.05\\ 
&$\alpha$          &&& $-$0.5 && $\pm$0.2  && $-$1.0  && $\pm$0.3\\
\hline
\end{tabular}
\end{center}
\vskip0.2cm {\small TABLE II: The critical exponents for the Gaussian
  distribution considered in this paper and for the binary
  distribution investigated in ref.\ \protect{\cite{Rieger}}, both for the
  constant field-strength/temperature ratio $h_r/T=0.35$} 
\vskip0.3cm

The first important observation is that the results obtained for a
Gaussian and a binary distribution do not differ significantly. In
particular the indications for a discontinuity in the magnetization
$\beta=0$ detected for the binary distribution in ref.\ \cite{Rieger}
are also present for the Gaussian distribution investigated in this
paper. This is a crucial point with regards to mean-field theory
\cite{Aharony}, where an essential difference between the two kinds of
distributions is predicted: In contrast to the continuous case with
nonvanishing weight at zero field strength the binary distribution is
expected to have a tricritical point, where the transition changes
from second to first order. Our conclusion in this paper is that in
three dimensions both distributions lead to identical results for the
investigated field strength.

Our estimate for the correlation length exponent $\nu$ has a rather
large errorbar, which is also the main source for the error in the
values for $\gamma$, $\overline{\gamma}$ and $\alpha$. One notes also
a discrepancy between the values of $\nu$ for the Gaussian and for the
binary distribution, but they remain compatible within the errorbars.
The Schwartz-Soffer inequality $\overline{\eta}\le2\eta$ seems to be
fulfilled as an equality, which supports the two-exponent scaling
scenario, as also found in \cite{Gofman}. An explicit calculation of
the amplitude ration in section III gives further evidence in
this respect.

The analysis of the probability distribution of the susceptibility
showed pronounced long tails as predicted by the scaling theories
\cite{Villain,Fisher}. The source for these long tails are rare
samples possessing a vanishing net-field and thus having a vanishing
expectation value for the magnetization. In fact these rare samples
provide the dominant contribution to the higher moments of the
distribution, which we checked explicitly. The analysis of the
probability distribution of the magnetization supports this picture
and shows that the probability of these rare samples indeed decays
like $L^{-\theta}$ with system size, where $\theta$ is an exponent
whose value turns out to be compatible with other predictions of the
scaling theory.

It turns out that the magnetization exponent $\beta$ is zero, which
means that the magnetization jumps discontinuously from zero to a
finite value at the critical temperature. Although this might be
called a first order phase transition a closer inspection of
the probability distribution of the magnetization revealed the absence
of a double-peak structure characterizing the phase-coexistence
usually present at a first order phase transition. 

Furthermore, at a first order phase transition the specific heat
usually diverges with the volume of the system \cite{Nauenberg}, but
the exponent $\alpha$ is negative, which means that the specific
heat does not diverge at $T_c$.  By means of birefringence
techniques Belanger et al.\ \cite{Belanger85} concluded from their
experiments on Fe$_{0.47}$Zn$_{0.53}$F$_{2}$ that $\alpha=0$ (i.e.\ a
logarithmic divergence of the specific heat). Only recently it was
shown \cite{Karszewiski} that the same kind of experiments on
Fe$_{0.85}$Mg$_{0.15}$Br$_{2}$ are better compatible with a cusp like
singularity of the specific heat and $\alpha=-1$, concurring with the
value reported in table II.  However, this value for $\alpha$ together
with the other estimates in table II would violate the modified
hyper-scaling relation (\ref{hyper}). Moreover, Schwartz
\cite{Schwartz} derived an exact inequality $2-\alpha\le\nu(d-2+\eta)$
and accepting $\eta\approx0.5$ (since this result has a much smaller
errorbar than $\nu$) it would imply $\nu\ge2$, which is a rather large
value. Indeed in a recent Migdal-Kadanoff\cite{Cao} such a large exponent 
was found: $\nu=2.25$, and also report
$\alpha=-1.37$ and $\beta=0.02$, consistent with table II.  However, a
value for $\alpha$ that is negative and large in modulus causes
serious difficulties with respect to the Rushbrooke relation 
(\ref{Rushbrooke}) and the more rigorous Rushbrooke inequality
$\alpha+2\beta+\gamma\ge2$ \cite{Rushbrooke}.

Let us conclude with the remark that this investigation on one hand
confirms the picture of the phase transition scenario found in ref.\
\cite{Rieger}. On the other hand it poses some serious puzzles 
concerning various scaling relations for the exponents that
still remain to be solved in the future. In addition
quite recently the possibility of an intermediate spin glass phase,
located between the paramagnetic and the ferromagnetic phase has been
discussed \cite{Mezard92,Mezard94}. It seems to us to be quite
important to clarify the consequences of the existence
such a phase, which is supposed to posses a {\it finite} correlation
length, for the nature of the transition considered here.

\section*{Acknowledgement}

I would like to thank A.\ P.\ Young for his encouragement and many
extremely helpful discussions.  The computations were done on the
Intel Paragon System from the Supercomputer center (HLRZ) at the
Forschungszentrum J\"ulich.  This work work was performed within the
SFB 341 K\"oln--Aachen--J\"ulich, supported by the DFG.

\vfill
\eject

\end{multicols}

\vfill
\eject

\eject

\eject

\eject

\end{document}